%% file: hdeccpv.tex
\documentclass[12pt]{article}
\usepackage{amsmath,amssymb,amsfonts,color,graphicx,cite,color}
\usepackage{colordvi,psfrag}
\usepackage{feynarts}
\input paperdef

\graphicspath{{plots/}}

\evensidemargin \oddsidemargin
\allowdisplaybreaks

\begin{document}
\thispagestyle{empty}

\def\thefootnote{\fnsymbol{footnote}}

\begin{flushright}
DCPT/07/150 \\
IPPP/07/75
\end{flushright}

\vspace{1cm}

\begin{center}

{\Large\sc {\bf Precise predictions for $h_a \to h_b h_c$
decays\\[0.4cm]
in the complex MSSM}}

\vspace{1cm}

{\sc
K.E.~Williams%
\footnote{email: K.E.Williams@durham.ac.uk}%
~and G.~Weiglein%
\footnote{email: Georg.Weiglein@durham.ac.uk}
}

\vspace*{.7cm}

{\sl
IPPP, University of Durham, Durham DH1~3LE, UK
}

\end{center}

\vspace*{0.1cm}

\begin{abstract}
\noindent
Complete one-loop results for the decay widths of neutral Higgs bosons
($h_a$) into lighter neutral Higgs bosons ($h_b,h_c$) are presented for
the MSSM with complex parameters. The results are obtained in the 
Feynman-diagrammatic
approach, taking into account the full dependence on the spectrum of
supersymmetric particles and all complex phases of the supersymmetric
parameters. The genuine triple-Higgs 
vertex contributions are supplemented with two-loop propagator-type 
corrections, yielding the currently most precise prediction for this class 
of processes. The genuine vertex corrections turn out to be very
important, yielding a large increase of the decay width compared to a
prediction based on the tree-level vertex.
The new results are used to analyse the impact of the
experimental limits from the LEP Higgs searches on the parameter space 
with a very light MSSM Higgs boson.
It is found that a significant part of the parameter space 
of the CPX benchmark scenario exists where 
channels involving the decay $h_2 \to h_1h_1$ have the highest search
sensitivity, and the existence of an unexcluded region
with $\MHe \approx 45 \gev$ is confirmed.
\end{abstract}

\def\thefootnote{\arabic{footnote}}
\setcounter{page}{0}
\setcounter{footnote}{0}

\newpage


\section{Introduction}

Higgs self-couplings, i.e.\ triple-Higgs couplings, $h_ah_bh_c$, and
quartic Higgs couplings, $h_ah_bh_ch_d$, are a crucial element of 
electroweak symmetry breaking via the Higgs mechanism. While the
prospects for a direct experimental determination of the quartic Higgs
coupling at present and future colliders are small (see e.g.\
\citere{LHCrev}),
probing the triple-Higgs coupling
will be one of the prime goals in the experimental programme for 
testing the Higgs mechanism. 
This coupling can be accessed via a precision measurement of the Higgs
production process $e^+e^- \to Zh_ah_a$ at the ILC~\cite{ilc} or
CLIC~\cite{clic},
and via Higgs cascade decays of the form $h_a \to h_bh_c$.
While Higgs cascade decays are obviously impossible in the
Standard Model (SM), they can play an important role in models with
extended Higgs sectors.

Besides the interest in Higgs cascade decays as a means to directly
probe Higgs self-couplings, a precise prediction for decays of this kind
is also important for phenomenological reasons. Where kinematically
possible the Higgs cascade decay modes can even be dominant and affect
Higgs phenomenology very significantly.

A well-known example of an extended Higgs sector is
the Minimal Supersymmetric Standard Model (MSSM), whose Higgs
sector comprises two scalar doublets that accommodate five physical Higgs
bosons. In
lowest order these are the light and heavy $\cp$-even $h$
and $H$, the $\cp$-odd $A$, and the charged Higgs bosons $H^\pm$.
Higher-order contributions yield large corrections to the masses
and couplings, and  also induce $\cp$-violation leading to
mixing between $h,H$ and $A$ in the case of general complex SUSY-breaking
parameters. The corresponding mass eigenstates are denoted as 
$h_1$, $h_2$, $h_3$. If the mixing between the three neutral mass
eigenstates is such that the coupling of the lightest Higgs boson to
gauge bosons is significantly suppressed, this state can be very light
without being in conflict with the exclusion bounds from the LEP Higgs
searches~\cite{LEPHiggsMSSM}. In this case the second-lightest Higgs boson, 
$h_2$, may predominantly decay into a pair of light Higgs bosons,
$h_2 \to h_1h_1$.

The results of the Higgs searches at LEP~\cite{LEPHiggsSM,LEPHiggsMSSM} have been 
interpreted in certain MSSM benchmark scenarios~\cite{benchmark,cpx}.
In the CPX scenario~\cite{cpx}, which involves large complex phases of
the trilinear coupling, $\At$, and
the gluino mass parameter, $M_3$,
the decay 
$h_2 \to h_1h_1$ is important in a significant part of the parameter
space. This decay mode leads to a quite complicated final state, 
corresponding to a
6-jet topology if $h_2$ is produced in association with an hadronically
decaying $Z$~boson or in association with $h_1$. 
In the analysis of the LEP Higgs searches within the CPX scenario an
unexcluded region for light $h_1$ and relatively small values of
$\tb$ (the ratio of the vacuum expectation values of the two Higgs doublets) 
remained, so that no firm lower bound on the mass of the lightest Higgs
boson of the MSSM could be set. The unexcluded parameter region with a
very light Higgs boson will also be difficult to cover with the Higgs
searches at the LHC~\cite{LHCrev,schumi,cpnsh} (see also \citere{ststh}
for a recent study).%
\footnote{
It should be noted that Higgs cascade decays are possible in
the MSSM also in the limit where complex phases are neglected. The 
decay $h \to A A$ occurs in a small parameter region with very light
$\MA$~\cite{svenhAA}, leading to small unexcluded  parameter
regions in the
$\MA$--$\tb$ plane from LEP Higgs
searches~\cite{LEPHiggsMSSM} (especially in the ``no-mixing''
scenario~\cite{benchmark}). For large values of $\MA$ also the decay 
$H \to h h$ can occur.
Higgs cascade decays into very light Higgs bosons can also be important in 
a considerable part of the parameter space of extensions of the MSSM, see 
e.g.\ \citere{nmssmhdec} for a discussion within the NMSSM.
}

In order to reliably determine which parameter regions of the MSSM with
a very light Higgs boson are unexcluded by the Higgs searches so
far and which regions will be accessible by Higgs searches in the
future, precise predictions for the Higgs cascade decays 
$h_a \to h_bh_c$ in the MSSM with complex parameters (cMSSM) are indispensable.
For propagator-type corrections the evaluations are quite advanced, 
and results incorporating the dominant one- and two-loop contributions
have been obtained within the Feynman-diagrammatic (FD)
approach~\cite{fd1,mhcMSSMlong,mhcpv2l} 
and the renormalisation-group (RG) improved effective
potential approach~\cite{rg}. The public codes
\fh~\cite{fhrandproc,feynhiggs,mhcMSSMlong,mhcpv2l},
based on the FD approach,
and \cpsh~\cite{cpsh}, based on the RG improved effective potential
approach, are available. For the genuine 
$h_ah_bh_c$ vertex contributions, on the other hand, so far 
only effective coupling approximations have been available in the cMSSM.

In this paper we obtain complete one-loop results within the FD approach
for the decays $h_a \to h_b h_c$ taking into account the
full dependence on all complex phases of the supersymmetric parameters. 
The mixing of the three neutral 
Higgs bosons among themselves and with the $Z$~boson and the unphysical
Goldstone-boson degree of freedom are consistently taken into account. 
We furthermore obtain complete one-loop results for the decays of 
neutral Higgs bosons into SM fermions in the cMSSM, $h_a \to f \bar f$.
The new one-loop results are combined with all existing higher-order
corrections in the FD approach, yielding in this way the currently most
precise predictions for the class of processes  $h_a \to h_b h_c$.
The results presented in this paper will be included in the code
\fh~\cite{fhrandproc,feynhiggs,mhcMSSMlong,mhcpv2l}.
We find that the genuine vertex corrections are very important for a 
reliable prediction of the Higgs cascade decays. The genuine vertex
corrections lead to a drastic change
compared to a prediction taking into account propagator-type corrections
only. We compare our full result with various approximations and
investigate the dependence on the complex phases. As an application of our 
improved theoretical predictions, we analyse to what extent the
previously unexcluded parameter region with a rather light Higgs boson
is compatible with the limits on topological cross sections obtained
from the LEP Higgs searches.


\section{The MSSM with complex parameters: notations and conventions}
\label{sec:cMSSM}

The MSSM, in its most general form, contains $\cp$-violating phases at tree 
level in the Higgs, slepton, squark, chargino/neutralino and gluino sectors. 
The gauge-boson, lepton and quark sectors do not contain extra phases (we 
assume in this paper a unit CKM matrix). In our calculation the full
dependence on all complex phases of the supersymmetric parameters
is taken into account. In the
following we briefly specify the
notations and conventions used in this paper and define the
parameters that are relevant for the discussion of the numerical
results.

\subsection{The squark sector}
The bilinear terms of the squarks in the Langrangian give rise to the
mass matrix
\begin{align}
M_{\sq} =
\begin{pmatrix}
        M_L^2 + \mq^2 + \MZ^2 \CZb (I_3^q - Q_q \sw^2) & \mq \; \Xq^* \\
        \mq \; \Xq    & M_{\tilde{q}_R}^2 + \mq^2 + \MZ^2 \CZb Q_q \sw^2
\end{pmatrix}  ,
\label{squarkmassmatrix}
\end{align}
where 
\BEA
\Xq &\equiv& A_q - \mu^* \{\CTb, \tb\} ,
\label{squarksoftSUSYbreaking}
\EEA
and the trilinear couplings $A_q$ and the Higgs-mixing parameter $\mu$ 
can be complex. This mass matrix needs to be diagonalised to get the 
tree-level physical states $\sqe,\sqz $,
\BE
\VL \sqe \\ \sqz \VR = \ML \ctq & 
\stq \\ -\stq^* & \ctq \MR\VL \sql \\ \sqr \VR ,
\end{equation}
where in our conventions $\ctq$ is real and $\stq$ is complex. 

\subsection{The neutral Higgs sector}
\subsubsection{Tree level}
We write the two Higgs doublets as
\begin{align}
\label{eq:higgsdoublets}
\cHe = \begin{pmatrix} H_{11} \\ H_{12} \end{pmatrix} &=
\begin{pmatrix} v_1 + \tfrac{1}{\sqrt{2}} (\phi_1-i \chi_1) \\
  -\phi^-_1 \end{pmatrix}, \notag \\ 
\cHz = \begin{pmatrix} H_{21} \\ H_{22} \end{pmatrix} &=
\begin{pmatrix} \phi^+_2 \\ v_2 + \tfrac{1}{\sqrt{2}} (\phi_2+i
  \chi_2) \end{pmatrix} ,
\end{align}
where $v_1$ and $v_2$ are the vacuum expectation values, and
$\tb \equiv v_2/v_1$. The conventions 
used here are the same as in \citere{mhcMSSMlong}. 
The MSSM Higgs sector is $\cp$-conserving at lowest order.
The tree-level neutral mass eigenstates $h,H,A$ and the unphysical
Goldstone-boson degree of freedom $G$ are related to the
$\cp$-even neutral fields $\phi_1,\phi_2$ and the $\cp$-odd neutral fields 
$\chi_1,\chi_2$ through a unitary matrix, 
\begin{align}
\begin{pmatrix} h \\ H \\ A \\ G \end{pmatrix} = \begin{pmatrix}
        - \sina & \cosa &                 0 &           0 \\
    \quad \cosa & \sina &                 0 &           0 \\
              0 &     0 &     - \sin \betan & \cos \betan \\
              0 &     0 & \quad \cos \betan & \sin \betan
  \end{pmatrix} 
\begin{pmatrix} \phi_1 \\ \phi_2 \\ \chi_1 \\ \chi_2 \end{pmatrix} .
\end{align}
In the renormalisation prescription that we have adopted, the parameter
$\tb$ receives a counterterm, while the mixing angle $\betan$ remains
unrenormalised. After the renormalisation has been carried out, one can
set $\betan = \be$.

\subsubsection{Higgs mass matrix in higher orders}
To find the loop-corrected neutral Higgs masses, 
the poles of the $3\times 3$ propagator matrix $\matr{\Delta}(p^2)$ 
in the $(h,H,A)$ basis need to be found (mixing with the Goldstone boson
$G$ and the $Z$~boson can be neglected in the propagator matrix since
the corresponding contributions are of sub-leading two-loop order, 
see the discussion in \citere{mhcMSSMlong}).
Determining the poles of the propagator matrix
is equivalent to finding the three solutions to 
\begin{equation}
\frac{1}{\left| \matr{\Delta}(p^2)\right|} = 0 .
\label{eq:findpoles}
\end{equation}
The propagator matrix is related to the $3\times3$ matrix of the 
irreducible renormalised
2-point vertex-functions $\matr{\hat{\Gamma}}_2(p^2)$ through
\begin{equation}
\left[-\matr{\Delta}(p^2)\right]^{-1} = 
\matr{\hat{\Gamma}}_2(p^2)=  i \left[p^2 \id - \matr{M}(p^2) \right] ,
\label{eq:gamma2}
\end{equation}
where
\BEA
  \matr{M}(p^2) &=
  \begin{pmatrix}
    \mh^2 - \ser{hh}(p^2) & - \ser{hH}(p^2) & - \ser{hA}(p^2) \\
    - \ser{hH}(p^2) & \mH^2 - \ser{HH}(p^2) & - \ser{HA}(p^2) \\
    - \ser{hA}(p^2) & - \ser{HA}(p^2) & \mA^2 - \ser{AA}(p^2)
\label{eq:massmatrix}
  \end{pmatrix}  . 
\EEA
Here $\mh$, $\mH$, $\mA$ are the lowest-order mass eigenvalues, and 
the $\ser{ij}$ ($i,j = h, H, A$) are the renormalised self-energies.
In general, the three solutions ${\cal M}_{h_a}^2$ (with $h_a=h_1,h_2,h_3$) 
are complex. They can be written as
\begin{equation}
{\cal M}_{h_a}^2 = M_{h_a}^2 - i \, M_{h_a} W_{h_a} ,
\end{equation}
where $M_{h_a}$ is the (loop-corrected) mass of the respective Higgs boson, 
$W_{h_a}$ is its width, and by convention
\begin{equation}
M_{h_1} \leq M_{h_2} \leq M_{h_3} .
\end{equation}

We calculated the self-energies using an expansion about $M^2_{h_a}$,
\BE
\ser{jk}({\cal M}_{h_a}^2) \approx
\ser{jk}(M_{h_a}^2)
+i\im \left[ {\cal M}_{h_a}^2 \right] \ser{jk}^{\prime}(M_{h_a}^2) ,
\label{se1order}
\end{equation}
with $j, k=h,H,A$ (as a check of our procedure, 
each time $\ser{jk}({\cal M}_{h_a}^2)$ was calculated, the next term in the 
expansion was also explicitly calculated, to ensure it is negligible). 
To find each solution to \refeq{eq:findpoles}, an iterative procedure was used.


\section{Complete one-loop calculation of Higgs-boson cascade decays
  and decays into SM fermions}

\label{sec:onel}

\begin{figure}[htb!]
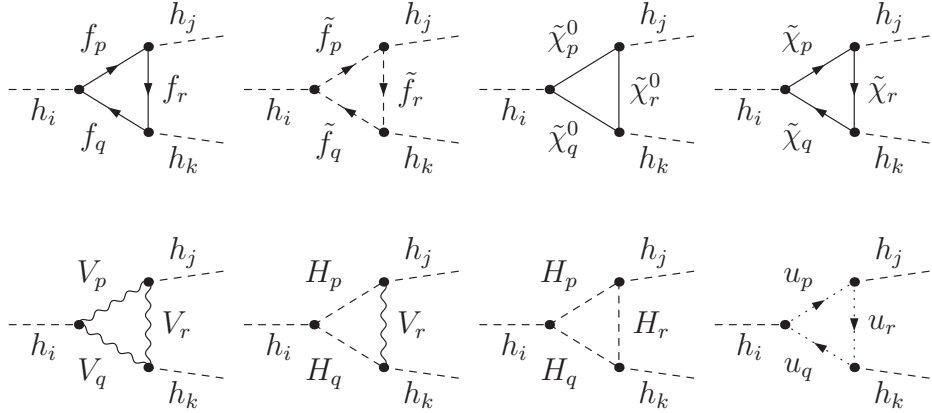

\begin{center}
\input hAhBhCdiag
\caption{Examples of generic diagrams (showing only one of the
topologies) contributing to the processes
$h_i \to h_j h_k$, where $h_i,h_j,h_k = h, H, A$.
Furthermore, 
$f$ are SM fermions, $\tilde{f}$ are their superparters, 
$\tilde{\chi}^0,\tilde{\chi}$ are neutralinos and charginos, 
$V$ are vector bosons, $H$ denote the neutral and charged Higgs bosons and 
the Goldstone bosons, $u$ are Faddeev--Popov ghost fields.
}
\label{fig:fd_hhh}
\end{center}
\end{figure}

We calculated the full 1PI (one-particle irreducible) one-loop vertex
contributions to the processes $h_a \rightarrow h_b h_c$, taking into
account all sectors of the MSSM and the complete dependence on the
complex phases of the supersymmetric parameters. Examples of generic
diagrams for one of the contributing topologies are shown in
\reffi{fig:fd_hhh}.

In order to obtain precise predictions for the branching ratios
$\br{(h_a \to h_b h_c)}$ it is important to calculate also the decay
widths of the Higgs bosons into SM fermions,
$h_a \to f \bar{f}$, at the same level of accuracy,
since the decay modes into $b \bar b$ and $\tau^+\tau^-$ are dominant over
large parts of the cMSSM parameter space. We therefore derived also
complete one-loop results for the processes $h_a \to f \bar{f}$ 
(including SM-type QED and, where appropriate, QCD corrections) for
arbitrary values of all complex phases of the supersymmetric parameters.
The partial widths for the other Higgs-boson decay modes have been taken
from the program
\fh~\cite{fhrandproc,feynhiggs,mhcMSSMlong,mhcpv2l}.

Our calculations have been carried out in the FD approach, making use of
the programs \fa~\cite{feynarts} and \fc~\cite{formcalc}.
Concerning the renormalisation, we use the same transformations and 
renormalisation conditions as in \citere{mhcMSSMlong}. We parameterise the
electric charge in the lowest-order decay amplitudes in terms of
$\al(\MZ)$, corresponding to the charge counterterm
\begin{equation}
\frac{\delta e}{e}= \frac{1}{2} \Pi_{\ga}(0) 
 - \frac{\sw}{\cw} \frac{\Si^{\rm T}_{\gamma Z}(0)}{\MZ^2}
 - \frac{1}{2} \De\al .
\end{equation}
Here $\Pi_{\ga}(0)$ is the photon vacuum polarisation, 
$\Si^{\rm T}$ denotes the transverse part of the self-energy, and
$\De\al=\De\al^{(5)}_{\textup{had}}+\De\al_{\textup{lept}}$ is the 
shift in the fine-structure constant arising from large logarithms of
light fermions. The other 
parameter renormalisations are listed in \citere{mhcMSSMlong}.
The fermion fields in the processes $h_a \to f \bar f$ are renormalised
on-shell. For the
renormalisation of the Higgs fields it is convenient to use a \drbar\
scheme as in \citere{mhcMSSMlong}, while the correct on-shell properties
of the S-matrix elements involving external Higgs fields are ensured by
the inclusion of finite wave-function normalisation factors. 

The wave-function normalisation factors are obtained from 
\BEA
Z_h=\frac{1}{
\left.\frac{\partial}{\partial p^2}\left(\frac{i}{\De_{hh}(p^2)}\right)\right|}
_{p^2={\cal M}_{h_a}^2}&
Z_H=\frac{1}{
\left.\frac{\partial}{\partial p^2}\left(\frac{i}{\De_{HH}(p^2)}\right)\right|}
_{p^2={\cal M}_{h_b}^2}&
Z_A=\frac{1}{
\left.\frac{\partial}{\partial p^2}\left(\frac{i}{\De_{AA}(p^2)}\right)\right|}
_{p^2={\cal M}_{h_c}^2}
\label{eq:Z1}\\
Z_{hH}=\left.\frac{\De_{hH}}{\De_{hh}}\right|_{p^2={\cal M}_{h_a}^2}&
Z_{Hh}=\left.\frac{\De_{hH}}{\De_{HH}}\right|_{p^2={\cal M}_{h_b}^2}&
Z_{Ah}=\left.\frac{\De_{hA}}{\De_{AA}}\right|_{p^2={\cal M}_{h_c}^2}\\
Z_{hA}=\left.\frac{\De_{hA}}{\De_{hh}}\right|_{p^2={\cal M}_{h_a}^2}&
Z_{HA}=\left.\frac{\De_{HA}}{\De_{HH}}\right|_{p^2={\cal M}_{h_b}^2}&
Z_{AH}=\left.\frac{\De_{HA}}{\De_{AA}}\right|_{p^2={\cal M}_{h_c}^2} ,
\label{eq:Z3}
\EEA
where the $\De_{hh}(p^2)$, $\De_{hH}(p^2)$ etc.\ are the elements of the
$3 \times 3$ Higgs propagator matrix (see \citere{mhcMSSMlong} for
expressions in terms of renormalised self-energies),
and $h_a,h_b,h_c$ are some permutation of $h_1,h_2,h_3$. 
For the evaluation of self-energies at complex momenta the expansion in
\refeq{se1order} is employed.

The wave-function normalisation factors can be expressed in terms of a 
(non-unitary) matrix $\matr{\hat Z}$~\cite{mhcMSSMlong}, 
\BEA
 \matr{\hat Z}  &=
  \begin{pmatrix}
    \sqrt{Z_h} & \sqrt{Z_h} Z_{hH} & \sqrt{ Z_h} Z_{hA} \\
    \sqrt{Z_H}Z_{Hh} & \sqrt{ Z_H} & \sqrt{ Z_H} Z_{HA} \\
    \sqrt{Z_A}Z_{Ah} & \sqrt{ Z_A} Z_{AH} & \sqrt{ Z_A}
  \end{pmatrix} , 
\EEA
where
\begin{align}
\begin{pmatrix} \hat\Ga_{h_a} \\ \hat\Ga_{h_b} \\ \hat\Ga_{h_c} \end{pmatrix} = \matr{\hat Z} \cdot
\begin{pmatrix} \hat\Ga_h \\ \hat\Ga_H \\\hat \Ga_A \end{pmatrix} .
\end{align}
Here $\hat\Ga_{h_a}$ is a one-particle irreducible n-point vertex-function 
(including loop corrections)
which involves a single external Higgs boson $h_a$.

The matrix $\matr{\hat Z}$ fulfills the conditions
\BEA
\label{eq:Zfactors1}
\lim_{p^2 \to {\cal M}_{h_a}^2}-\frac{i}{p^2-{\cal M}_{h_a}^2}\left(\matr{\hat
Z}\cdot\matr{\hat \Gamma}_2\cdot\matr{\hat Z}^{T}\right)_{hh}&=&1\\
\lim_{p^2 \to {\cal M}_{h_b}^2}-\frac{i}{p^2-{\cal M}_{h_b}^2}\left(\matr{\hat
Z}\cdot\matr{\hat \Gamma}_2\cdot\matr{\hat Z}^{T}\right)_{HH}&=&1\\
\lim_{p^2 \to {\cal M}_{h_c}^2}-\frac{i}{p^2-{\cal M}_{h_c}^2}\left(\matr{\hat
Z}\cdot\matr{\hat \Gamma}_2\cdot\matr{\hat Z}^{T}\right)_{AA}&=&1 ,
\EEA
where $\matr{\hat \Gamma}_2$ has been introduced in \refeq{eq:gamma2}.
We choose in the following $h_a=h_1$, $h_b=h_2$ and $h_c=h_3$. It should
be noted that this choice is purely a convention. Other choices would give 
the same physical results. This fact can most easily be seen if the
wave-function normalisation factors are defined as in
\refeqs{eq:Z1}--(\ref{eq:Z3}), i.e., with the Higgs-boson
self-energies evaluated at the complex pole. In this way the evaluation
of the masses and the wave-function normalisation factors is treated on
the same footing. The definition of the wave-function normalisation
factors adopted in \refeqs{eq:Z1}--(\ref{eq:Z3}) differs slightly from
the one in \citeres{mhcMSSMlong,karinaproc}, where the wave-function
normalisation factors were defined at the real part of the complex pole
(furthermore, in \citeres{mhcMSSMlong,karinaproc} the real parts of the
diagonal wave-function normalisation factors $Z_h$, $Z_H$, $Z_A$
were taken). The two definitions of the wave-function normalisation
factors differ by contributions from imaginary parts that are formally
of sub-leading two-loop order. It turns out that the inclusion of the
imaginary parts improves the numerical stability of the results in
certain parameter regions, while otherwise these contributions are
completely negligible (see also \citere{lcws07higgs}).

In a complete one-loop calculation of Higgs decay processes also mixing
contributions between the states $h,H,A$ and the Goldstone and $Z$ bosons
have to be taken into account (we denote these reducible contributions
as $\Gamma_{h_ih_jh_k}^{\rm G,Z \, mix}$). We treat these mixing contributions
strictly at one-loop level in order to ensure the cancellation of
unphysical poles (while our prescription for the wave function
normalisation factors described above automatically incorporates leading
reducible higher-order contributions). Since the mixing self-energy involving 
$G$ and $Z$ is already a one-loop contribution, in a strict one-loop
expansion the
$\Gamma_{h_ih_jh_k}^{\rm G,Z \, mix}$ are evaluated at the (unrotated)
tree-level masses $m^2_{h_i}$. Accordingly, our results can be written
as (summation over repeated indices is understood)
\BEA
\Gamma_{h_ah_bh_c}^{\rm full}&=&
\matr{\hat{Z}}_{ck} \matr{\hat{Z}}_{bj} \matr{\hat{Z}}_{ai}
\left[\Gamma_{h_ih_jh_k}^{\rm 1PI}\left(M^2_{h_a},M^2_{h_b},M^2_{h_c}\right)+
\Gamma_{h_ih_jh_k}^{\rm G,Z \, mix}\left(m^2_{h_i},m^2_{h_j},m^2_{h_k}\right)
\right]\\
\Gamma_{h_af\bar{f}}^{\rm full}&=&
\matr{\hat{Z}}_{ai}\left[\Gamma_{h_if\bar{f}}^{\rm 1PI}\left(M^2_{h_a}\right)+
\Gamma_{h_if\bar{f}}^{\rm G,Z \, mix}\left(m^2_{h_i}\right)\right] ,
\EEA
where as before $h_i,h_j,h_k = h,H,A$.
The numerical impact of the mixing contributions with the Goldstone and
$Z$~bosons on the results presented in this paper turned out to be
small.

In our numerical analysis below we will compare our full result with the 
contribution obtained from the $t,\tilde{t}$ sector in the
approximation where the gauge couplings are neglected and the diagrams
are evaluated at zero external momenta. We refer to this approximation
as the ``Yukawa Approximation'', which is expected to yield the leading
one-loop contribution if $\tb$ is not too large.
In this approximation, the counterterm contributions in the renormalised 
vertex all vanish.


\section{Combination with higher-order results}

\label{sec:combi}

We have combined our new one-loop result with the most up-to-date
higher-order propagator-type contributions in the FD approach. This has
been done by supplementing the one-loop self-energies in 
\refeqs{eq:massmatrix} and (\ref{eq:Z1})--(\ref{eq:Z3})
with the two-loop self-energies obtained from the program 
\fh~\cite{fhrandproc,feynhiggs,mhcMSSMlong,mhcpv2l}.
In \fh\ the \order{\alt\als} corrections are incorporated including the
full phase dependence at the two-loop level, while other two-loop
contributions so far are only known in the limit of vanishing complex
phases~\cite{rMSSMcorr}. In our numerical analysis below 
we will restrict to those higher-order
contributions for which the phase dependence at the two-loop is
explicitly known, i.e.\ we do not include the residual contributions for
which a result exists only in the MSSM with real parameters.
The calculation of the decay width 
$\Gamma (h_a\rightarrow b\bar{b})$ 
furthermore contains resummed SUSY-QCD corrections, including the full
phase dependence.

By supplementing our complete one-loop results for the processes 
$h_a \to h_b h_c$ and $h_a \to f \bar f$ with the state-of-the-art
higher-order propagator-type corrections we obtain the currently 
most precise predictions for the $h_a \to h_b h_c$ decay widths and 
branching ratios in the MSSM with complex parameters. It should be noted
that in the special case where the complex phases are put to zero 
our results also provide improved predictions for the Higgs cascade
decays occurring in the $\cp$-conserving case, i.e.\ $h \to AA$ and 
$H \to hh$. The numerical impact of the latter will be discussed
elsewhere.

In our numerical analysis we also investigated the impact of using 
loop-corrected Higgs masses and couplings (rather than the tree-level
values) within loop diagrams, which is formally a higher-order effect.
The numerical impact on our results turned out to be negligible.


\section{Implementation of exclusion bounds from the LEP Higgs searches}

\label{sec:higgsbounds}

A precise prediction for the process $h_2 \to h_1 h_1$ is particularly
important for investigating the exclusion bounds from the Higgs searches at
LEP~\cite{LEPHiggsSM,LEPHiggsMSSM} within the MSSM with complex parameters. The LEP
analysis in the CPX benchmark scenario~\cite{cpx} resulted in an
unexcluded parameter region for relatively small $\tb$ where the
lightest neutral Higgs boson is very light but has strongly suppressed
couplings to gauge bosons. The second-lightest Higgs boson, on the other
hand, may be within the kinematic reach of LEP in this region but has a
large branching ratio into a pair of lightest Higgs bosons, i.e.\
$h_2 \to h_1 h_1$.

In our numerical analysis below we analyse the impact of our new result 
on the LEP exclusion bounds in the cMSSM. This is done by comparing
the cMSSM predictions with the topological cross section limits given in
\citeres{LEPHiggsSM,LEPHiggsMSSM,philip}. The LEP limits on the various cross
sections~\cite{LEPHiggsSM,LEPHiggsMSSM} have been implemented into the code 
{\em HiggsBounds}~\cite{higgsbounds} (for applications of preliminary
versions of {\em HiggsBounds}, see \citere{ollirefs}). 
In order to obtain a correct
statistical interpretation of the overall exclusion limit at the 95\%
C.L., the first step is to determine, for each parameter point, which
one of the various channels has the highest statistical sensitivity 
for setting an exclusion limit~\cite{philip}. We then compare the
theoretical prediction for this particular channel
with the topological cross
section limit determined at LEP for this channel.  We neglect, in this
analysis, the theoretical
uncertainties from unknown higher-order corrections --- see
\citere{higgsbounds} for a discussion of
this issue. The predictions of the topological cross sections have been 
obtained using the wave-function normalisation factors defined in 
\refse{sec:onel}.

While the topological cross section
limits are very convenient for testing a wide class of models and are not
restricted to specific parameter values, it should be kept in mind that
the dedicated analyses carried out in \citere{LEPHiggsMSSM} for certain
benchmark scenarios have a higher statistical sensitivity than the
limits obtained from the topological cross sections. This is in
particular the case in parameter regions where two (or more) channels
have a similar statistical sensitivity, since the method based on the 
topological cross section limits allows one only to use one channel at a
time.


\section{Numerical results}

In our numerical analysis we use the parameter values of the CPX
benchmark scenario~\cite{cpx}, adapted to the latest experimental
central value of the top-quark mass~\cite{mt1709} and using an on-shell
value for the absolute value of the trilinear couplings 
$\At$ and $\Ab$ 
that is somewhat shifted compared to the 
\drbar\ value specified in \citere{cpx} (see \citere{bse} for a
discussion in the MSSM with real parameters). Specifically, if not
indicated differently we use the
following parameters (the lowest-order Higgs-sector parameters 
$\tb$ and $\MHp$ are varied; in our analysis we include $\tb$ values up
to 40 and $\MHp$ values up to $1000 \gev$.)
\begin{align}
& \msusy = 500 \gev, \; |\At| = |\Ab| = 900 \gev, \;
  \mu = 2000 \gev, \; \mgl \equiv |M_3| = 1000 \gev , \non \\
& \mt = 170.9 \gev, \; M_2 = 200 \gev ,
\label{eq:cpx1}
\end{align}
and the complex phases of the trilinear couplings $\At$, $\Ab$
and the gluino mass parameter $M_3$ are set to
\begin{equation}
\phiat = \frac{\pi}{2}, \quad  \phiab = \frac{\pi}{2}, 
\quad \phigl = \frac{\pi}{2} .
\label{eq:cpx2}
\end{equation}
In \refeq{eq:cpx1}
$\msusy$ denotes the diagonal soft SUSY-breaking parameters in the
sfermion mass matrices that are chosen to be equal to each other,
$\msusy = M_L = M_{\tilde{q}_R}$,
see \refeq{squarkmassmatrix}.

\subsection{Results for the $h_2\rightarrow h_1h_1$ decay width}

\begin{figure}
\begin{tabular}{cc}
\includegraphics[height=0.5\linewidth,angle=0]{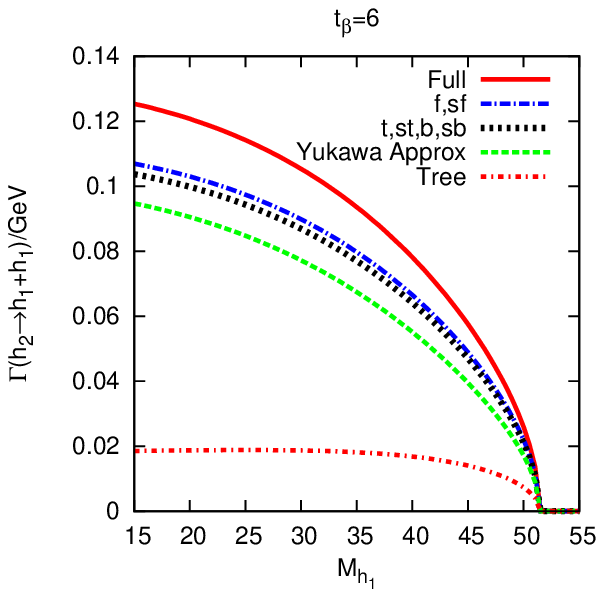}&
\includegraphics[height=0.5\linewidth,angle=0]{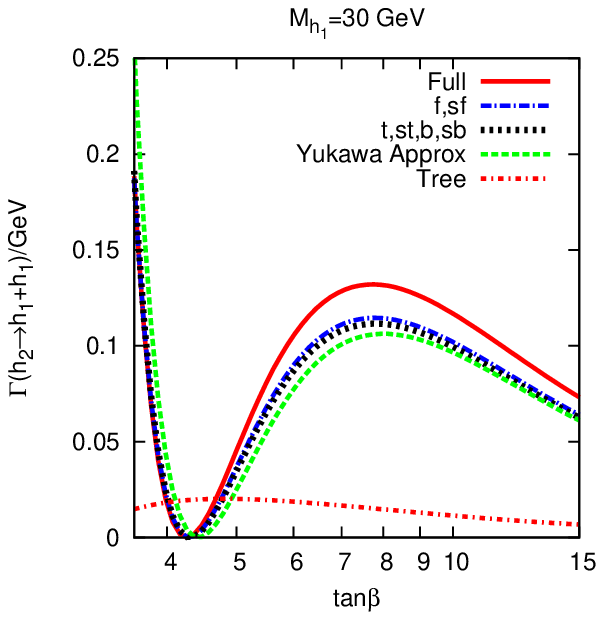}\\
(a)&(b)\\
\end{tabular}
\caption{Full result for $\Ga(h_2 \to h_1h_1)$ compared with various
approximations, see text.
(a) $\Ga(h_2 \to h_1h_1)$ vs.\ $\MHe$ for $\tb = 6$ ($\MHp$ is
varied; all other parameters are set to the CPX values given in
\refeqs{eq:cpx1}, (\ref{eq:cpx2})). 
(b) $\Ga(h_2 \to h_1h_1)$ as function of $\tb$ for $\MHe = 30 \gev$
($\MHp$ is adjusted to ensure $\MHe = 30 \gev$).
\label{linegraphs}
}
\end{figure}

\reffi{linegraphs} shows the relative effect of different contributions to 
the $h_2\rightarrow h_1h_1$ decay width in the area of the cMSSM parameter
space which is particularly relevant to an investigation of the
unexcluded regions 
in the LEP Higgs searches~\cite{LEPHiggsMSSM}. 
In \reffi{linegraphs} (a) 
the full result for the decay width and various approximations 
are plotted against $\MHe$ while keeping $\tb=6$. 
All the results plotted include the higher-order corrected wave-function
normalisation factors as described in \refses{sec:onel} and
\ref{sec:combi}. They differ only in the genuine contributions to the 
$h_2h_1h_1$ vertex. One can see that the full result (denoted as
`Full') differs drastically from the case where only wave-function
normalisation factors but no genuine one-loop vertex contributions are
taken into account (`Tree'). 
The inclusion of the genuine vertex
corrections that have been evaluated in this paper can increase the 
decay width by more than a factor of six in this example
(for values of $\MHe$ sufficiently below the kinematic limit of 
$\MHe = 0.5 \MHz$ where the decay width goes to zero). The Yukawa
approximation agrees much better with the full result, giving rise to
deviations of up to $\sim 30\%$. 
Using the full contribution from the $t,\tilde{t},b,\tilde{b}$ sector 
(`t, st, b, sb') and from all three generations of fermions and
sfermions (`f, sf') yields a prediction that deviates from the full
result by up to 20\%. 

In \reffi{linegraphs} (b), the decay width is plotted against $\tb$ whilst 
adjusting $\MHp$ such that $\MHe = 30 \gev$. The meaning of the various
lines is the same as in \reffi{linegraphs} (a). The result where the
genuine vertex corrections are neglected (`Tree') shows a completely
different qualitative behaviour and differs drastically from the full
result. The pronounced dependence on $\tb$, giving rise in particular to 
a region where $\Ga(h_2 \to h_1h_1) \approx 0$ for $\tb \approx 4.3$,
is due to the fact that the Yukawa vertex corrections (which are dominant) 
to the matrix element change sign when $\tb$ is varied. As the decay width 
depends on the absolute value squared of the matrix element, 
the region where the Yukawa vertex corrections are close to zero 
corresponds to a minimum of the decay width. The deviations between the
full result and the contribution from only the fermion and sfermion
sector reach about 15\% in this example.

\begin{figure}
\begin{center}
\includegraphics[height=0.5\linewidth,angle=0]{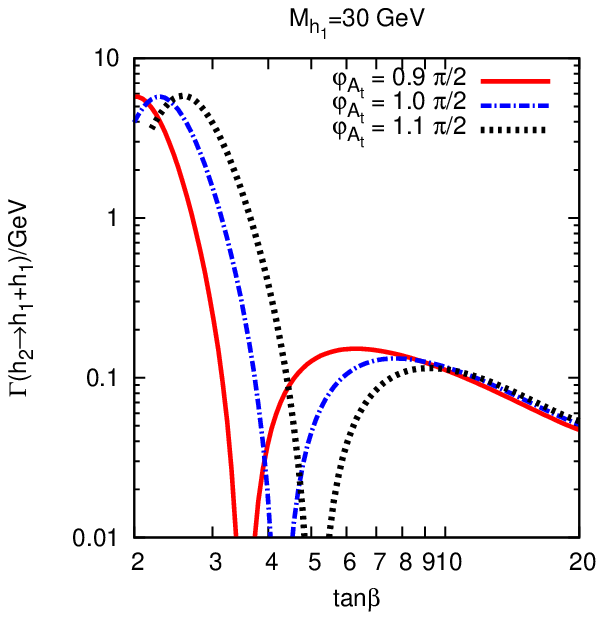}
\end{center}
\caption{$\Ga(h_2 \to h_1h_1)$ as function of $\tb$ for $\MHe = 30 \gev$
and different values of $\phiat$
($\MHp$ is adjusted to ensure $\MHe = 30 \gev$). 
\label{varyArgAt}
}
\end{figure}

In \reffi{varyArgAt} the full result for $\Ga(h_2 \to h_1h_1)$ is shown
as a function of $\tb$ for $\MHe = 30 \gev$ (adjusting $\MHp$
accordingly) and different values of $\phiat$. The dependence on
$\phiat$ is very pronounced, leading in particular to a relative shift
of the curves in $\tb$. As a consequence, comparing the results for 
$\Ga(h_2 \to h_1h_1)$ for different values of $\phiat$ at fixed values
of $\tb$ can yield dramatic effects. Thus, a thorough treatment of the
phase dependence is indispensable for a meaningful theoretical
prediction of Higgs cascade decays in this parameter region.

\subsection{Analysis of exclusions bounds from the LEP Higgs searches}

\begin{figure}[htb!]
\begin{center}
\includegraphics[height=0.70\linewidth,angle=0]{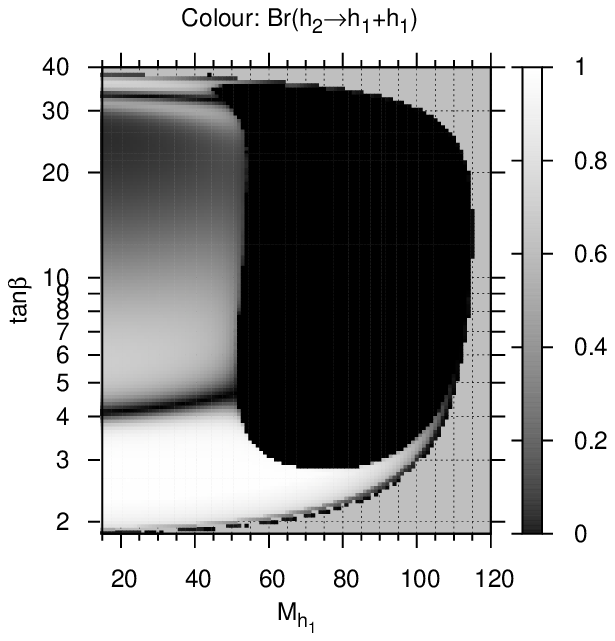}
\end{center}
\vspace{-2em}
\caption{The branching ratio $\br{(h_2\to h_1h_1)}$ in the $\MHe$--$\tb$
plane of the CPX scenario.
\label{MHpTBplots1}
}
\end{figure}

\begin{figure}[htb!]
\begin{tabular}{cc}
\includegraphics[height=0.45\linewidth,angle=0]{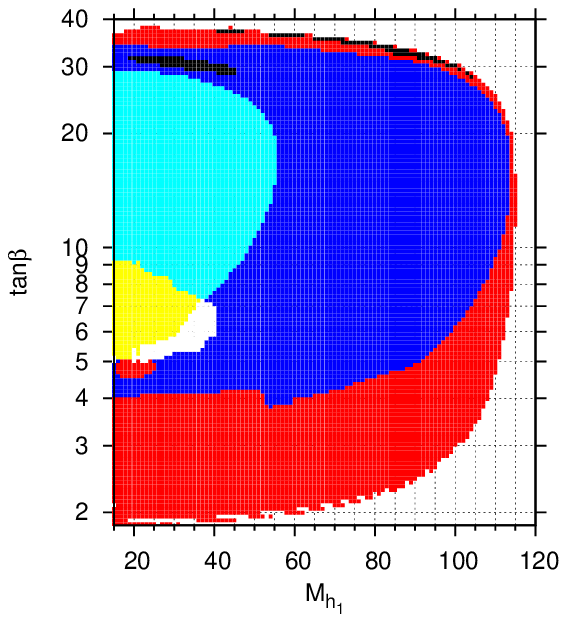}&
\includegraphics[height=0.45\linewidth,angle=0]{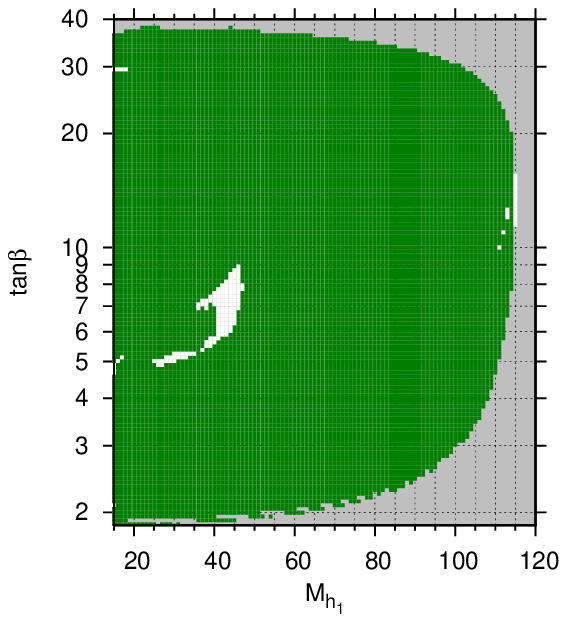}\\[-.5em]
(a)&(b)
\end{tabular}
\caption{Coverage of the LEP Higgs searches in the $\MHe$--$\tb$ plane
of the CPX scenario. Plot (a) shows the channels that are 
predicted to have the highest statistical sensitivity for setting an
exclusion limit.
The colour codings
in are: red $({\color{red}\blacksquare})=$ $(h_1 Z)\to(b \bar b Z)$,
blue $({\color{blue}\blacksquare})=$ $(h_2  Z)\to(b \bar b Z)$,
white $(\square)=$ $(h_2  Z)\to(h_1 h_1 Z)\to(b \bar b b \bar b Z)$,
cyan $({\color{cyan}\blacksquare})=$ $(h_2 h_1)\to(b \bar b b \bar b)$,
yellow $({\color{yellow}\blacksquare})=$ $(h_2 h_1)\to(h_1 h_1 h_1)\to(b \bar b b \bar b b \bar b)$,
black $(\blacksquare)=$ other channels.
Plot (b) shows the parameter regions excluded 
at the 95\% C.L.\
by the topological cross section limits obtained at LEP~\cite{LEPHiggsSM, LEPHiggsMSSM}.
The colour codings are:
green (dark grey) = LEP excluded, white = LEP allowed.
\label{MHpTBplots2}
}
\end{figure}

In \reffis{MHpTBplots1} and \ref{MHpTBplots2} the $\MHe$--$\tb$ parameter 
space of the CPX scenario is analysed. \reffi{MHpTBplots1} shows the 
branching ratio of the Higgs
cascade decay, $\br{(h_2\to h_1h_1)}$. It can be seen that, over a large
part of the parameter space where the decay $h_2\to h_1h_1$ is
kinematically possible, it is actually the dominant decay channel. The
branching ratio is particularly large for low and moderate values of
$\tb$. In the region where $\tb \approx 4$--5 the Yukawa contribution to
the matrix element for the $h_2\to h_1h_1$ decay changes sign
and causes a sharp drop in the $h_2\to h_1h_1$ branching ratio,
as was already observed in \reffi{linegraphs}. A similar behaviour 
occurs also in
the region of $\tb \approx 35$. 

Plot (a) of \reffi{MHpTBplots2} indicates which channel has the highest
statistical sensitivity and therefore which channel will be used to set
an exclusion limit in different 
regions of the parameter space. As explained in
\refse{sec:higgsbounds}, this information is needed for
an interpretation of the topological cross section limits 
obtained at LEP~\cite{LEPHiggsSM,LEPHiggsMSSM} as 95\% C.L.\ 
excluded regions in the 
$\MHe$--$\tb$ parameter space. 
One can see in the figure that the
channels $e^+e^- \to h_2  Z\to  h_1 h_1 Z \to b \bar b b \bar b Z$
and $e^+e^- \to h_2 h_1 \to h_1 h_1 h_1 \to b \bar b b \bar b b \bar b$
have the highest search sensitivity in a region with small $\MHe$ and
moderate values of $\tb$, $5 \lsim \tb \lsim 9$.
For small $\MHe$ and somewhat higher $\tb$ the channel 
$e^+e^- \to (h_2 h_1)\to(b \bar b b \bar b)$ has the highest search
sensitivity. It should be noted that all channels involving the decay of
the $h_2$ boson in the region of small $\MHe$ are strongly influenced by
the $\Ga(h_2 \to h_1h_1)$ decay width, either directly in the case of
the channels involving the Higgs cascade decay, or indirectly through
the branching ratio of the $h_2$. 
The parameter region where $\Ga(h_2 \to h_1h_1)$
is important coincides with the
region of the CPX scenario that could not be excluded at the 95\% C.L.\ 
in the analysis of the four LEP collaborations~\cite{LEPHiggsMSSM}.

In plot (b) of \reffi{MHpTBplots2} we have combined our new theoretical
predictions (containing the complete one-loop result for the genuine
vertex corrections and higher-order corrections, as described above)
with the topological cross section limits obtained at LEP.
We find an unexcluded region at $\MHe \approx 45 \gev$ and moderate
$\tb$ where channels involving the decay $h_2 \to h_1h_1$ play an
important role. Thus, our analysis, based on the most
up-to-date theory prediction for the $h_2 \to h_1h_1$ channel, confirms
the `hole' in the LEP coverage observed in \citere{LEPHiggsMSSM}
(see in particular Fig.~19 of \citere{LEPHiggsMSSM}). 

It should be noted, on the other hand, that the results shown in 
plot (b) of \reffi{MHpTBplots2} differ in several respects from the 
results presented in \citere{LEPHiggsMSSM}. As discussed above, 
near to borders between areas where different search topologies
are predicted to have the highest exclusion power our 
analysis has less statistical sensitivity than the benchmark
studies of \citere{LEPHiggsMSSM}. 
A further difference
is the input value of the top-quark mass. While
we are using the latest experimental central value of 
$\mt = 170.9 \gev$~\cite{mt1709}, most of the analysis of 
\citere{LEPHiggsMSSM} was
done for $\mt = 174.3 \gev$. We have explicitly checked that (as expected)
the unexcluded region is significantly increased if we use 
$\mt = 174.3\gev$ instead. Concerning differences in the theoretical 
predictions, the analysis of \citere{LEPHiggsMSSM} was based on the two
codes {\em FeynHiggs2.0}, an earlier version of the 
\fh\ program~\cite{feynhiggs}, 
and {\em CPH}~\cite{cpx}, a predecessor of the program \cpsh~\cite{cpsh}.
For each scan point, in \citere{LEPHiggsMSSM} the results from 
{\em CPH} and {\em FeynHiggs2.0} (for the decay width 
$\Ga(h_2 \to h_1h_1)$ the {\em CPH \/} formula was used in both codes) were
compared with each other, and the result yielding the more conservative
exclusion bound was retained. Our theoretical prediction for 
$\Ga(h_2 \to h_1h_1)$ based on a complete diagrammatic one-loop result
of the genuine vertex contribution goes significantly beyond the
effective coupling approximation used in \citere{LEPHiggsMSSM}.
Furthermore, the \order{\alt\als} propagator-type corrections obtained 
in the Feynman-diagrammatic approach for arbitrary complex 
phases~\cite{mhcpv2l} were not yet available when the analysis of 
\citere{LEPHiggsMSSM} was carried out.

\begin{figure}
\begin{tabular}{ccc}
\includegraphics[height=0.3\linewidth,angle=0]{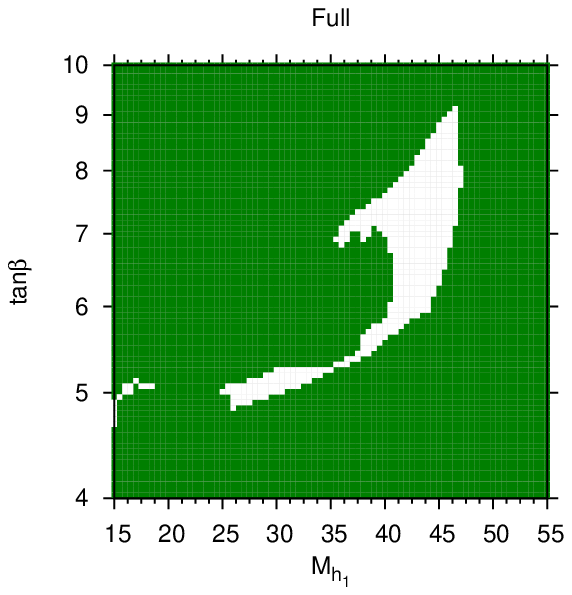}&
\includegraphics[height=0.3\linewidth,angle=0]{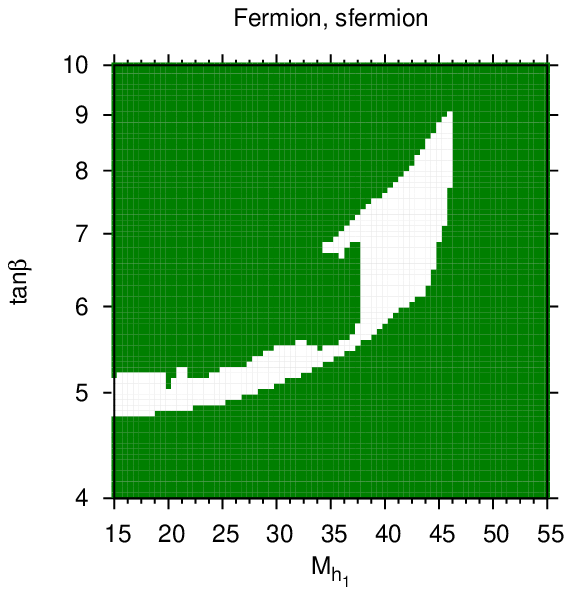}&
\includegraphics[height=0.3\linewidth,angle=0]{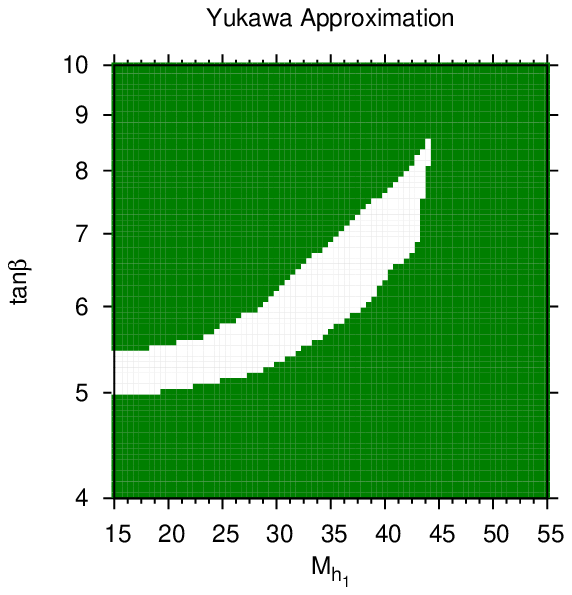}\\
(a)&(b)&(c)
\end{tabular}
\caption{Impact of the genuine vertex corrections on the `hole' in the 
LEP coverage for 
$\MHe \approx 45 \gev$ and $\tb \approx 7$. Plot (a) shows the full
result (detailed view of plot (b) of \reffi{MHpTBplots2}). Plot (b)
shows the result for the case where only contributions from SM fermions
and their superpartners are taken into account in the genuine vertex 
corrections. Plot (c) shows the result where the Yukawa Approximation
has been used for the genuine vertex corrections.
The colour codings are:
green (dark grey) = LEP excluded, white = LEP allowed.
\label{approxLEP}
}
\end{figure}

In \reffi{approxLEP} we focus on the uncovered parameter region at 
$\MHe \approx 45 \gev$ and $\tb \approx 7$ and compare our complete
one-loop result for the genuine triple-Higgs vertex corrections with 
two approximations. While plot (a) shows the full result (i.e., it is a 
detailed view of plot (b) of \reffi{MHpTBplots2}), in plot (b) the
genuine vertex corrections are approximated by the contributions from
fermions and sfermions only, and in plot (c) the Yukawa Approximation
has been used for the genuine vertex corrections. In all three plots the
wave function normalisation factors and all other decay widths are the
same (calculated as described above). The differences between the three
plots in \reffi{approxLEP} are therefore entirely caused by the genuine
vertex corrections to the Higgs cascade decays. 
While all three graphs show an unexcluded region at
$\MHe \approx 45 \gev$ and $\tb \approx 7$, the shape of this region
changes quite significantly.
In particular, the analysis based on the full result gives rise to a
considerably larger unexcluded region around $\MHe \approx 45 \gev$ and
$\tb \approx 7$ compared to the Yukawa Approximation (while the
Yukawa Approximation gives rise to a larger unexcluded region for
smaller $\MHe$). As expected from
\reffi{linegraphs}, the inclusion of all SM fermion and sfermion loop
contributions yields a better approximation of the full result.

\section{Conclusions}

We have obtained, within the MSSM with complex parameters,
complete one-loop results for the classes of decay processes in which a 
heavier neutral Higgs boson decays into two lighter ones and for those
in which a neutral Higgs boson decays into a pair of SM fermions. 
Our results take into account all sectors of the MSSM and incorporate the
full dependence on all complex phases of the supersymmetric parameters 
and the external momenta. The new one-loop results have been
supplemented with state-of-the-art propagator-type corrections 
containing higher-order contributions obtained in the FD approach
(taken from the program \fh), 
yielding in this way the currently most precise predictions for these
processes within the MSSM with complex parameters. 

We find that the genuine vertex contributions to the triple-Higgs vertex 
are numerically very important. Their inclusion
changes the predictions for the decay widths very drastically compared 
to the case of an effective coupling approximation based only on the
wave function normalisation factors of the external Higgs bosons.
Including genuine vertex corrections in a simple Yukawa Approximation 
yields a prediction for the decay width that is much closer to the full
result, but we still find deviations of up to 30\% in the numerical
examples investigated in this paper (using the CPX scenario). 
Even contributions beyond the
fermion/sfermion sector can have a sizable impact on the decay widths.
We have furthermore found that the dependence of the decay width on 
the complex phase $\phiat$ is very pronounced, emphasizing the need 
for a thorough treatment of the effects of complex phases.

Based on our improved theoretical predictions we have analysed 
the impact of the limits on topological cross sections obtained from the
LEP Higgs searches on the parameter space with a very light Higgs boson
within the MSSM with complex parameters. We find that, over a large
part of the parameter space where the decay $h_2\to h_1h_1$ is
kinematically possible, it is the dominant decay channel. The
corresponding search channels
$e^+e^- \to h_2  Z\to  h_1 h_1 Z \to b \bar b b \bar b Z$
and $e^+e^- \to h_2 h_1 \to h_1 h_1 h_1 \to b \bar b b \bar b b \bar b$
have the highest sensitivity for setting an exclusion limit 
in a region with small $\MHe$ and
moderate values of $\tb$. We find that a parameter region
with $\MHe \approx 45 \gev$ and $\tb \approx 7$
remains unexcluded by the limits on topological cross sections
obtained from the LEP Higgs searches, confirming the results of the four
LEP collaborations achieved in a dedicated analysis of the CPX benchmark
scenario. A precise theory prediction for the $h_2\to h_1h_1$ channel is
crucial for mapping out the unexcluded parameter region
(it should be noted in this context that in the parameter region of the
CPX scenario also formally sub-leading two-loop corrections can have a
sizable numerical impact). We find that
the shape of the unexcluded region is significantly modified if the full
result for the vertex corrections is replaced by approximations.

The results presented in this paper will be included in the public
code \fh. It would be interesting to compare our results with the other
public code for evaluating Higgs masses and decay widths in the MSSM
with complex parameters, \cpsh. This comparison is affected, however,
not only by the genuine triple-Higgs vertex corrections that are the 
main focus of the present paper, but also by differences in the
propagator-type corrections used in the two codes.
Furthermore, a 
meaningful comparison between \fh\ and  \cpsh\ requires a translation 
between the on-shell input parameters used in \fh\ and the \drbar\ input 
parameters used in \cpsh. Such a detailed comparison is beyond the scope
of the present paper. We will address this issue in a forthcoming
publication.


\subsection*{Acknowledgements}

We thank P.~Bechtle, O.~Brein, T.~Hahn, S.~Heinemeyer, W.~Hollik,
S.~Palmer, A.~Read, H.~Rzehak and D.~St\"ockinger for numerous helpful 
discussions. 
Collaboration with P.~Bechtle, O.~Brein and S.~Heinemeyer
on the implementation of the 
bounds from the LEP Higgs searches into the code {\em HiggsBounds\/} 
is gratefully acknowledged.
Work supported in part by the European Community's Marie-Curie Research
Training Network under contract MRTN-CT-2006-035505
`Tools and Precision Calculations for Physics Discoveries at Colliders'.



\end{document}

%% file: hAhBhCdiag.tex
\unitlength=1bp%

\begin{feynartspicture}(432,204)(4,2.3)

\FADiagram{}
\FAProp(0.,10.)(6.5,10.)(0.,){/ScalarDash}{0}
\FALabel(3.25,9.18)[t]{$h_i$}
\FAProp(20.,15.)(13.,14.)(0.,){/ScalarDash}{0}
\FALabel(16.3162,15.3069)[b]{$h_j$}
\FAProp(20.,5.)(13.,6.)(0.,){/ScalarDash}{0}
\FALabel(16.3162,4.69307)[t]{$h_k$}
\FAProp(6.5,10.)(13.,14.)(0.,){/Straight}{1}
\FALabel(9.20801,13.1807)[br]{$f_p$}
\FAProp(6.5,10.)(13.,6.)(0.,){/Straight}{-1}
\FALabel(9.20801,6.81927)[tr]{$f_q$}
\FAProp(13.,14.)(13.,6.)(0.,){/Straight}{1}
\FALabel(14.274,10.)[l]{$f_r$}
\FAVert(6.5,10.){0}
\FAVert(13.,14.){0}
\FAVert(13.,6.){0}

\FADiagram{}
\FAProp(0.,10.)(6.5,10.)(0.,){/ScalarDash}{0}
\FALabel(3.25,9.18)[t]{$h_i$}
\FAProp(20.,15.)(13.,14.)(0.,){/ScalarDash}{0}
\FALabel(16.3162,15.3069)[b]{$h_j$}
\FAProp(20.,5.)(13.,6.)(0.,){/ScalarDash}{0}
\FALabel(16.3162,4.69307)[t]{$h_k$}
\FAProp(6.5,10.)(13.,14.)(0.,){/ScalarDash}{1}
\FALabel(9.20801,13.1807)[br]{$\tilde{f}_p$}
\FAProp(6.5,10.)(13.,6.)(0.,){/ScalarDash}{-1}
\FALabel(9.20801,6.81927)[tr]{$\tilde{f}_q$}
\FAProp(13.,14.)(13.,6.)(0.,){/ScalarDash}{1}
\FALabel(14.274,10.)[l]{$\tilde{f}_r$}
\FAVert(6.5,10.){0}
\FAVert(13.,14.){0}
\FAVert(13.,6.){0}

\FADiagram{}
\FAProp(0.,10.)(6.5,10.)(0.,){/ScalarDash}{0}
\FALabel(3.25,9.18)[t]{$h_i$}
\FAProp(20.,15.)(13.,14.)(0.,){/ScalarDash}{0}
\FALabel(16.3162,15.3069)[b]{$h_j$}
\FAProp(20.,5.)(13.,6.)(0.,){/ScalarDash}{0}
\FALabel(16.3162,4.69307)[t]{$h_k$}
\FAProp(6.5,10.)(13.,14.)(0.,){/Straight}{0}
\FALabel(9.33903,12.9678)[br]{$\tilde \chi_p^0$}
\FAProp(6.5,10.)(13.,6.)(0.,){/Straight}{0}
\FALabel(9.33903,7.03218)[tr]{$\tilde \chi_q^0$}
\FAProp(13.,14.)(13.,6.)(0.,){/Straight}{0}
\FALabel(14.024,10.)[l]{$\tilde \chi_r^0$}
\FAVert(6.5,10.){0}
\FAVert(13.,14.){0}
\FAVert(13.,6.){0}

\FADiagram{}
\FAProp(0.,10.)(6.5,10.)(0.,){/ScalarDash}{0}
\FALabel(3.25,9.18)[t]{$h_i$}
\FAProp(20.,15.)(13.,14.)(0.,){/ScalarDash}{0}
\FALabel(16.3162,15.3069)[b]{$h_j$}
\FAProp(20.,5.)(13.,6.)(0.,){/ScalarDash}{0}
\FALabel(16.3162,4.69307)[t]{$h_k$}
\FAProp(6.5,10.)(13.,14.)(0.,){/Straight}{1}
\FALabel(9.20801,13.1807)[br]{$\tilde \chi_p$}
\FAProp(6.5,10.)(13.,6.)(0.,){/Straight}{-1}
\FALabel(9.20801,6.81927)[tr]{$\tilde \chi_q$}
\FAProp(13.,14.)(13.,6.)(0.,){/Straight}{1}
\FALabel(14.274,10.)[l]{$\tilde \chi_r$}
\FAVert(6.5,10.){0}
\FAVert(13.,14.){0}
\FAVert(13.,6.){0}

\FADiagram{}
\FAProp(0.,10.)(6.5,10.)(0.,){/ScalarDash}{0}
\FALabel(3.25,9.18)[t]{$h_i$}
\FAProp(20.,15.)(13.,14.)(0.,){/ScalarDash}{0}
\FALabel(16.3162,15.3069)[b]{$h_j$}
\FAProp(20.,5.)(13.,6.)(0.,){/ScalarDash}{0}
\FALabel(16.3162,4.69307)[t]{$h_k$}
\FAProp(6.5,10.)(13.,14.)(0.,){/Sine}{0}
\FALabel(9.20801,13.1807)[br]{$V_p$}
\FAProp(6.5,10.)(13.,6.)(0.,){/Sine}{0}
\FALabel(9.20801,6.81927)[tr]{$V_q$}
\FAProp(13.,14.)(13.,6.)(0.,){/Sine}{0}
\FALabel(14.274,10.)[l]{$V_r$}
\FAVert(6.5,10.){0}
\FAVert(13.,14.){0}
\FAVert(13.,6.){0}

\FADiagram{}
\FAProp(0.,10.)(6.5,10.)(0.,){/ScalarDash}{0}
\FALabel(3.25,9.18)[t]{$h_i$}
\FAProp(20.,15.)(13.,14.)(0.,){/ScalarDash}{0}
\FALabel(16.3162,15.3069)[b]{$h_j$}
\FAProp(20.,5.)(13.,6.)(0.,){/ScalarDash}{0}
\FALabel(16.3162,4.69307)[t]{$h_k$}
\FAProp(6.5,10.)(13.,14.)(0.,){/ScalarDash}{0}
\FALabel(9.20801,13.1807)[br]{$H_p$}
\FAProp(6.5,10.)(13.,6.)(0.,){/ScalarDash}{0}
\FALabel(9.20801,6.81927)[tr]{$H_q$}
\FAProp(13.,14.)(13.,6.)(0.,){/Sine}{0}
\FALabel(14.274,10.)[l]{$V_r$}
\FAVert(6.5,10.){0}
\FAVert(13.,14.){0}
\FAVert(13.,6.){0}

\FADiagram{}

\FAProp(0.,10.)(6.5,10.)(0.,){/ScalarDash}{0}
\FALabel(3.25,9.18)[t]{$h_i$}
\FAProp(20.,15.)(13.,14.)(0.,){/ScalarDash}{0}
\FALabel(16.3162,15.3069)[b]{$h_j$}
\FAProp(20.,5.)(13.,6.)(0.,){/ScalarDash}{0}
\FALabel(16.3162,4.69307)[t]{$h_k$}
\FAProp(6.5,10.)(13.,14.)(0.,){/ScalarDash}{0}
\FALabel(9.20801,13.1807)[br]{$H_p$}
\FAProp(6.5,10.)(13.,6.)(0.,){/ScalarDash}{0}
\FALabel(9.20801,6.81927)[tr]{$H_q$}
\FAProp(13.,14.)(13.,6.)(0.,){/ScalarDash}{0}
\FALabel(14.274,10.)[l]{$H_r$}
\FAVert(6.5,10.){0}
\FAVert(13.,14.){0}
\FAVert(13.,6.){0}

\FADiagram{}\FAProp(0.,10.)(6.5,10.)(0.,){/ScalarDash}{0}
\FALabel(3.25,9.18)[t]{$h_i$}
\FAProp(20.,15.)(13.,14.)(0.,){/ScalarDash}{0}
\FALabel(16.3162,15.3069)[b]{$h_j$}
\FAProp(20.,5.)(13.,6.)(0.,){/ScalarDash}{0}
\FALabel(16.3162,4.69307)[t]{$h_k$}
\FAProp(6.5,10.)(13.,14.)(0.,){/GhostDash}{1}
\FALabel(9.20801,13.1807)[br]{$u_p$}
\FAProp(6.5,10.)(13.,6.)(0.,){/GhostDash}{-1}
\FALabel(9.20801,6.81927)[tr]{$u_q$}
\FAProp(13.,14.)(13.,6.)(0.,){/GhostDash}{1}
\FALabel(14.274,10.)[l]{$u_r$}
\FAVert(6.5,10.){0}
\FAVert(13.,14.){0}
\FAVert(13.,6.){0}
\end{feynartspicture}